\documentclass[twocolumn]{aastex6}
\usepackage{natbib,amssymb,amsmath,graphicx}
\usepackage{hyperref}
\newcommand\tna{\,\tablenotemark{a}}
\newcommand\tnb{\,\tablenotemark{b}}

\newcommand\cdB{\mbox{CD-35~2722~B }}

\newcommand\cd{\mbox{CD-35~2722 }}

\begin{document}

\title{L-band Spectroscopy with Magellan-AO/Clio2: First Results on Young
Low-Mass Companions} 

\author{Jordan M. Stone\altaffilmark{1}}
\author{Josh Eisner\altaffilmark{1}}
\author{Andy Skemer\altaffilmark{1,2,5}}
\author{Katie M. Morzinski\altaffilmark{1,6}}
\author{Laird Close\altaffilmark{1}}
\author{Jared Males\altaffilmark{1}}
\author{Timothy J. Rodigas\altaffilmark{3,5}}
\author{Phil Hinz\altaffilmark{1}}
\author{Alfio Puglisi\altaffilmark{4}}
\altaffiltext{1}{Steward Observatory,
University of Arizona,
933 N. Cherry Ave,
Tucson, AZ 85721-0065 USA} 
\altaffiltext{2}{Department of Astronomy and Astrophysics, 
University of California, Santa Cruz, 
1156 High St, 
Santa Cruz, CA 95064, USA}
\altaffiltext{3}{Department of Terrestrial Magnetism,
Carnegie Institution of Washington,
5241 Broad Branch Road NW,
Washington, DC 20015, USA}
\altaffiltext{4}{INAF-Osservatorio Astrofisico di Arcetri, I-050125, Firenze, Italy}
\altaffiltext{5}{Hubble Fellow}
\altaffiltext{6}{Sagan Fellow}

\begin{abstract} 
L-band spectroscopy is a powerful probe of cool low-gravity atmospheres: The P,
Q, and R branch fundamental transitions of methane near 3.3~$\mu$m provide
a sensitive probe of carbon chemistry; cloud thickness modifies the spectral
slope across the band; and H$_{3}^{+}$ opacity can be used to detect aurorae.
Many directly imaged gas-giant companions to nearby young stars exhibit L-band
fluxes distinct from the field population of brown dwarfs at the same effective
temperature.  Here we describe commissioning the L-band spectroscopic mode of
Clio2, the 1-5 $\mu$m instrument behind the Magellan adaptive-optics system.
We use this system to measure L-band spectra of directly imaged companions. Our
spectra are generally consistent with the parameters derived from previous
near-infrared spectra for these late M to early L type objects. Therefore,
deviations from the field sequence are constrained to occur below 1500 K. This
range includes the L-T transition for field objects and suggests that
observed discrepancies are due to differences in cloud structure and
CO/CH$_{4}$ chemistry. 
\end{abstract}

\section{Introduction} 
Low-gravity atmospheres, like those of young gas-giant planets, display
distinct characteristics compared to their higher-gravity counterparts, even at
the same effective temperatures
\citep{Martin96,Marois08,Bowler10,Currie11,Barman11a,Skemer12,Marley12}. Both
cloudiness and dis-equilibrium chemistry are enhanced as a result of the
prevailing low pressure in the atmospheres of the young low-mass objects:
clouds because the photosphere is beneath the temperature-pressure location
where refractory elements begin to condense to liquid and solid phases; and
dis-equilibrium chemistry because the longer chemical timescales implied by the
low pressure (and density) cannot keep up with vertical mixing timescales that
deliver warm material from deeper layers \citep[e.g.,][]{Hubeny07, Barman11b,
Marley12}.

As demonstrated in the cases of 2M~1207~b \citep{Chauvin04} and HR~8799~b, c,
and d \citep{Marois08}, dis-equilibrium chemistry, vertical mixing, and
cloudiness are important factors in understanding the observed features of cool
young/low-gravity atmospheres \citep{Skemer11,Barman11a,Barman11b}.  Clouds and
dis-equilibrium CH$_{4}$/CO chemistry have a large effect on L-band spectra
because the P-,Q-,and R-branch methane bandheads trace the abundance of
CH$_{4}$, and cloudiness affects the general slope of the spectrum
\citep{Yamamura2010,Skemer14}. L-band measurements
augmenting near-IR (Y, J, H, and K band) spectra can also help to break
degeneracies (e.g., between metalicity and gravity) when performing fits to
model atmospheres \citep{Stephens09,Skemer15}.

Atmospheric properties are crucial for understanding the nature and origin of
directly imaged companions.  The HR~8799 system contains four directly imaged
companions with a non-hierarchical orbital architecture that closely resembles
a planetary system, albeit much more massive then typical
\citep{Marois08,Marois10}. Yet with uncertain companion masses
($\sim10~M_{\mathrm{Jup}}$) it is possible (though unlikely) that at least some
of the companions exceed the $\sim12~M_{\mathrm{Jup}}$ minimum mass for
deuterium burning\footnote{The maximum mass for an extrasolar planet according
to the current IAU definition, which suggests that planets have masses below
the deuterium burning limit and are gravitationally bound to a more massive
primary. See \citet{Chabrier14} for a discussion of some of the issues with
this definition.}.  PSO~J318.5-22, a $\sim6~M_{\mathrm{Jup}}$ isolated member
of the $\beta$~Pic moving group, occupies the same locus of color-magnitude
space as HR~8799~b, c, d, e, and 2M 1207 b, but this object is not a planet
because it was likely born in isolation \citep{Liu13}.

PSO~J318.5-22 and other low-mass brown dwarfs in the field appear to form
through an extension of the star-formation process to very low masses
\citep{Luhman12}. Similarly, 2M~1207~b likely formed as an extension of the
binary-star formation process \citep{Lodato05}. The orbital architecture of the
HR~8799 system suggests a distinct formation mechanism, such as core-accretion
or gravitational-instability in a circum-stellar disk \citep{Kratter10}. Thus
it appears nature is capable of producing fundamentally different objects, with
distinct formation processes that nevertheless have overlapping ``planetary"
masses.

Spectroscopy constrains the formation mechanisms of directly imaged companions
by revealing the composition of their atmospheres \citep{Konopacky13,Barman15}.
If companion atmospheres gain a significant fraction of their mass via the
accretion of solids in a disk, then they should show increased metalicity. The
core-accretion process for gas-giant planet formation can produce metal
enhanced planetary atmospheres compared to the host star \citep{Pollack96}.
Gravitational fragmentation of a massive protoplanetary disk can also result in
a metal enriched atmosphere \citep{Boss97, Boley11}.  Furthermore, because of
the radial temperature gradient in protoplanetary disks, certain chemical
species will be removed from the gas phase at radii corresponding to their
freeze-out temperatures. This will result in distinct elemental ratios in the
planetary atmosphere which may be diagnostic of the radial position at which
a given planet formed in a disk \citep[e.g.,][]{Oberg11}. On the other hand, if
a companion is produced via gravitational fragmentation of the initial dense
molecular cloud core, no large compositional differences are expected. The
utility of spectroscopy for discriminating between formation modes relies on
compositional differences being manifest in the atmospheres of directly imaged
companions and may be affected by different assumptions regarding how material
from deep within an object can be transported to the outer envelope
\citep[e.g.,][]{Thiabaud15}.

The Magellan adaptive-optics system \citep[MagAO;][]{Close13}, with its
adaptive secondary mirror, provides low-background L-band images at the
diffraction limit of the 6.5 m Magellan Clay telescope (providing
a point-spread function with a width of 121 milliarcseconds). The mid-infrared
camera, Clio2 \citep[like Clio,][but with an upgraded HAWAII-1
array]{Sivanandam06,Hinz10} that operates behind the MagAO system, is equipped
with a prism that yields increasing spectral resolution from $R\sim50$--$300$
across the L-band. Given the sensitive high-spatial resolution capabilities of
MagAO, Clio2 spectroscopy is a powerful tool for providing the L-band
spectroscopic constraints necessary for a complete understanding of directly
imaged low-mass companion atmospheres. MagAO/Clio2 complements
ongoing near-IR surveys for the detection and spectral characterization of
new companions, such as the Gemini Planet Imager Exoplanet Survey
\citep{Macintosh15}, and surveys with the SPHERE instrument at the VLT
\citep{Beuzit08}, which are not sensitive in the L-band.

In this paper, we establish the L-band spectroscopic mode of MagAO/Clio2 and
report first results on low-mass directly imaged companions. In Section
\ref{ObsSec} we describe our sample of targets and summarize near-IR constraints on
the nature of their atmospheres. We also describe our data collection and
reduction approach for the new mode. In Section \ref{ResultsSec} we summarize
our L-band spectroscopic results for the sample. Finally, in Section
\ref{DiscSec} we discuss the implications of our findings.

\section{Observations and Reduction} \label{ObsSec}

Data for this work were collected during two observing runs in 2014. The first
was from 2014 April 5--6, and the second was from 2014 November 13--15.
A summary of our observing logs is presented in Table \ref{Obslog}. Our data
collection and calibration approach is described below.

\subsection{Target Selection and Parameters from Literature}
\begin{deluxetable*}{llcrrll}
\tabletypesize{\footnotesize}
\tablecolumns{8}
\tablewidth{0pt}
\tablecaption{Observing log}
\tablehead{
\colhead{Target} &
\colhead{Date} & 
\colhead{Plate Scale} & 
\colhead{Tot. Int.}&
\colhead{R\tna}&
\colhead{guidestar mag\tnb}&
\colhead{Comments}\\
\colhead{} &
\colhead{} & 
\colhead{[mas/pix]} & 
\colhead{[s]}&
\colhead{}&
\colhead{}&
\colhead{}}
\startdata
\cdB & 2014-04-07&15.9& 280& 50--200&m$_{I}=$9&\\ 
TWA~5~B      & 2014-04-07&15.9&2240& 50--200&m$_{I}=$9.1&\\ 
HD~138575    & 2014-04-07&15.9& 840& 50--200&m$_{V}=$7&A0V telluric calibrator\\ 
\\
HD~32007     & 2014-11-14&25.7& 540&150--300&m$_{V}=$8.5&B9 telluric calibrator\\ 
AB~Pic~b     & 2014-11-14&25.7&3690&150--300&m$_{R}=$9&\\ 
$\eta$~Tel~B & 2014-11-15&25.7& 720&150--300&m$_{R}=$5&\\ 
2M~0103(AB)~b& 2014-11-15&25.7&1080&150--300&m$_{I}=$12.9&guidestar is $\sim0.2''$ binary\\ 
HD~10553     & 2014-11-15&25.7& 360&150--300&m$_{V}=$6.6&A3V telluric calibrator\\ 
\enddata \label{Obslog}
\tablenotetext{a}{The prism provides increasing spectral resolution with
wavelength across the L-band. The realized resolution is a function of the
AO-performance because the PSF is smaller than the slit.}
\tablenotetext{b}{For our companion sources we used the primary stars as the
AO-guidestar. The wavefront sensor uses a non-standard filter covering the
R and I bands.}
\end{deluxetable*}
\begin{deluxetable*}{lllllrlccl}
\tabletypesize{\footnotesize}
\tablecolumns{10}
\tablewidth{0pt}
\tablecaption{Summary of target information from the literature}
\tablehead{
                                    &
                                    &
                                    &
                                    &
\multicolumn{2}{c}{NIR template fit\tna}&
                                    &
                                    &
\\
\colhead{Name}                      &
\colhead{Dist.}                     & 
\colhead{Sep.}                      & 
\colhead{Sp. type}                  & 
\colhead{ $T_{\mathrm{eff}}$ }         &
\colhead{log($g$)}                  &
\colhead{Assoc.}                    &
\colhead{Est. Mass}                  &
\colhead{Est. Age}                 &
\colhead{references}\\
                              &
\colhead{[pc]}                & 
\colhead{[$''$]}              & 
                              & 
\colhead{[K]}                 &
\colhead{[cm s$^{-2}$]}          &
                              &
\colhead{[$M_{\mathrm{Jup}}$]}          &
\colhead{[Myr]}                         &
}
\startdata
$\eta$~Tel~B   & 48 & 4.2 & M7-9 & 2600$\pm$100        &4$\pm$0.5   &$\beta$~Pic&$\sim$30 &$\sim$12&5,12--14\\ 
TWA~5~B        & 50 & 2   & M8.5$\pm$0.5 & 2500$\pm$100        &4$\pm$0.5   &TW Hya     &$\sim$20 &$\sim$10&5,8--11\\
\cdB   & 21 & 3.1 & L3-L4           & 1700-1900           &4.5$\pm$0.5 &AB~Dor     &$\sim$13 &$\sim$50&2,3,15\\ 
AB~Pic~b       & 46 & 5.5 & L0$\pm$1     & 1800$^{+100}_{-200}$&4.5$\pm$0.5 &Tuc-Hor    &$\sim$12 &$\sim$30&4--7\\ 
2M~0103(AB)~b  & 47 & 1.7 & L            &---                  &---         &Tuc-Hor    &$\sim$13 &$\sim$30&1,7\\ 
\enddata
\tablenotetext{a}{The near-IR spectrum of CD-35 27722 B was fit with an
AMES-Dusty model atmosphere. For all the other sources best fit parameters from
the BT-Settl 2012 grid are reported}
\tablerefs{(1) \citet{Delorme13}; (2) \citet{Wahhaj11}; (3) \citet{Zuckerman04}; (4) \citet{Chauvin05}; 
          (5) \citet{Bonnefoy14}; (6) \citet{Bonnefoy10}; (7) \citet{Torres00}; (8) \citet{Lowrance99}; (9) \citet{Neuhauser00}; 
          (10) \citet{Weinberger13}; (11) \citet{Webb99}; (12) \citet{Lowrance00}; (13) \citet{Neuhauser11}; 
          (14) \citet{Zuckerman01}; (15) \citet{Allers13}}
\label{targSumTable}
\end{deluxetable*}

Our target list, summarized in Table \ref{targSumTable}, was selected to
include nearby young stars with low-mass companions visible from Las Campanas Observatory
(declination $\lesssim20^{\circ}$). Given the sensitivity of the MagAO
wavefront sensor and of Clio2, we also selected targets to have guidestar
I-band magnitudes $\lesssim13$ and companion L-band magnitudes $\lesssim14$ to
facilitate obtaining adequate signal to noise in less than one night of observing. We
focused on targets with well characterized near-IR spectra for direct
comparison to our results. One of our targets, 2MASS~J01033563-5515561(AB)b
(hereafter 2M~0103(AB)~b), which orbits a 0.2$''$ M6-M6 binary-star system,
does not have near-IR spectra available in the literature. We added this target
to bolster our low-mass sample and to test the performance of MagAO while
guiding on a close nearly equal-magnitude binary.

\subsection{Instrumental Setup} 

The MagAO system is built around a 585 actuator adaptive secondary mirror
that minimizes the number of optical elements necessary for correcting the
blurring effects of Earth's atmosphere. Compared to alternative implementations
that position an adaptive element at a re-imaged pupil plane, adaptive-secondary
systems provide higher throughput and lower thermal background
\citep{Lloyd-Hart00}. The system provides correction for up to 300 modes at up
to 989 Hz. The pyramid wavefront sensor used by MagAO facilitates easy
adjustments both in the number of modes corrected and the loop speed to help
keep the AO loop locked on faint guidestars or in periods of poor seeing. 

The prism in Clio2 provides increasing spectral resolution with wavelength. At
J band the typical resolution is R$\sim8$ and at M band it is R$\sim500$. We
used a 260 milliarcsecond slit to minimize the bright mid-infrared sky emission
in our spectra. For observations with the PSF smaller than the slit, the size
of the PSF is the relevant parameter for determining the spectral resolution.
Thus, our observed spectral resolution changes with AO performance and
wavelength.  During the two runs reported here, we realized R=50--200 and
R=150--300 across the L-band. 

Due to the very different dispersions in the near- and mid-IR, there are 
different timescales for detector saturation in each band.  In order to reduce
practical complications related to saturation and cross-talk on the Clio2
detector, we observed through a blocking filter that only passes light from
$\sim2.8\mu\mathrm{m}$ -- $\sim4.2\mu\mathrm{m}$.  

Two camera lenses provide Clio2 with two plate scales, a coarse mode with 27.5
milliarcseconds pixel$^{-1}$, and a fine mode with 15.9 milliarcseconds
pixel$^{-1}$ \citep{Morzinski15}. As part of our effort to define best
practices for the use of the MagAO/Clio2 spectroscopy mode, we used a different
plate scale for each of our runs. This also accommodated the observation of
wider binaries. The optics used to set the plate scale modify the position
of the spectral trace on the Clio2 detector. The fine mode situates the trace
on the left side of the detector, and the coarse mode positions the trace on
the right side, which has more high dark current pixels and produces noisier
spectra as a result.

\subsection{Data Acquisition}
We used the fine plate scale and observed both TWA~5 and \cd during our
first run. For our second run, we used the coarse plate scale and observed
$\eta$~Tel~B, AB~Pic~b, and 2M~0103(AB)~b. We saved frames differently for each
run, saving single images per nod position during the first run and multiple
images per nod during the second run. This difference slightly changes our
reduction procedure for each dataset, as discussed below.

For each object, we aligned the binary position angle of our targets to be
parallel to the Clio2 slit so that we could collect spectra of both primary and
companion sources simultaneously. This orientation allows us to use the primary
star as a telluric calibrator for the companion. Because the slit position and
angle are not repeatable, and because Clio2 does not have a slit viewer, we
developed the following alignment procedure.

First, after acquiring our targets with the slit and prism out of the beam, we
noted the position of the primary star and the position angle of the
companion. Next, we introduced the slit and nudged the telescope horizontally
until the primary star could be seen in the center. We then measured the angle
of the slit on the detector and calculated the necessary slit rotation offset to
align the target PA with the slit.  The calculated rotation offset was
implemented in small steps, $\lesssim5^{\circ}$, without opening the AO loop.
By keeping the AO loop closed on the primary star while we offset the rotator
to align our targets, we ensured that the star remained on the rotation axis,
stationary in the slit.

Once both objects were aligned in the slit, we introduced the prism.  In order
to track variable background sky emission, we nodded our targets every 1--3
mins. When we could, we nodded along the slit so that we were always
integrating on the target. At some rotation angles, when the entire weight of
the wavefront sensor apparatus rested on just one motor, we had to nod off the
slit because the motors were insufficiently powered to provide the desired nod
vector. Currently both the previous weaker 1.6 Amp motors have been replaced by
3.0 Amp motors to hold the full weight.

On 2014 November 15, we observed $\eta$~Tel during twilight because the source
was setting. Since the sky was bright in the optical, we were unable to perform
the normal telescope collimation steps. We ran the AO loop closed with only
10-modes for five minutes in order to offload as much focus to the telescope as
possible, and then increased the number of modes to begin observing. Twilight
background light on the wavefront sensor still limited the accuracy of AO
corrections and poor AO-performance was realized for this source.

\subsection{Preliminary Reduction} \label{extractSec}

Our data reduction process began by subtracting nod pairs. For the data from
April this was simply an A-B and a B-A subtraction, where A and B represent our
two nod positions. For the data from November, when we saved 6 images per nod
position, we did $\mathrm{A}_{i}-\mathrm{median(B)}$ and
$\mathrm{B}_{i}-\mathrm{median(A)}$ , $i$=1--6. We then rotated each image to
produce horizontal spectral traces.  For the November data, we also applied
a vertical shift to align each of the 6 images per nod, and co-added them.
Typical rotations were $\sim1.3^{\circ}$ and typical vertical offsets were
$\lesssim 0.1$ pixels.

We created a master spectral trace for each object by combining the traces from
each nod sequence. For our April observations we combined the traces using
a simple mean. For our November observations, since we saved multiple images
per nod position, we were able to measure the image variance in each pixel in
each nod, so we derived a weight-image for each nod inversely proportional to
the image variance before combining. For all of our sources we also
bootstrapped the stacking procedure, randomly selecting individual frames (with
replacement), before stacking.  This provides a variance image for use in
determining optimal weights for spectral extraction (see below), and the raw
data for bootstrapping the extraction, wavelength calibration, and telluric
correction steps necessary for generating errorbars for our spectra.

\begin{figure}
\epsscale{0.7}
\plotone{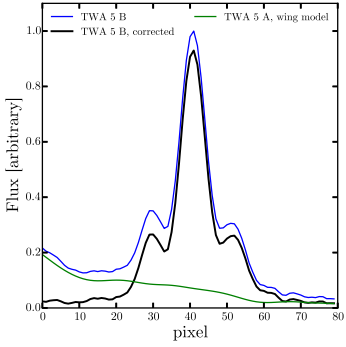}
\caption{A slice through the spatial dimension of the trace of TWA5 B. Light
from the unresolved binary host influences the left shoulder. We removed the
influence of the primary star using a model constructed by scaling the columns
of a telluric calibrator trace (see text).\label{twaProfFig}} 
\end{figure}

For close companions, spectra may be contaminated by the wings of the primary
if the primary flux at the position of the companion is larger than the
background noise.  TWA~5~B and 2M~0103(AB)~b are the closest in our sample,
requiring attention to this effect. To remove the light of the primary from the
spectrum of the secondary, we created a model of the primary star trace using
a high signal-to-noise image of a telluric calibrator spectrum observed just
after observing the target.  We rotated and shifted the telluric trace to
overlap the trace of the primary.  We then scaled each column of the telluric
image so that the pixels in the core and first Airy ring best-fit the
corresponding pixels in the image of the primary. This provided us with a model
of the trace of the primary without the secondary.  We smoothed this model with
a circular gaussian kernel with $\sigma=4$ pixels, and then subtracted it from
the image before extracting the companion spectrum. In Figure \ref{twaProfFig}
we show the average spatial profile of TWA~5~B before and after correction.

For the 2M~0103 system, we achieved much lower signal-to-noise than for TWA~5.
Therefore our model trace was also noisy. Since the magnitude of the
correction for 2M~0103(AB)~b was much smaller than the noise, we decided not to
subtract a noisy model from the images before extraction.

\subsection{Spectral Extraction}

We used the wavelength dependent spatial profiles of primary stars and our
bootstrap generated variance maps to define optimal weights \citep{Horne86} for
extraction. We found that including the first Airy ring in the spatial profile
was important to avoid fringing in our spectra because spectral traces exhibit
a fringe pattern where the core and Airy rings oscillate out of phase. For
2M~0103(AB)~b, the circumbinary source, the primary spatial profile was
corrupted by light from the nearby secondary. In this case we used the spatial
profile of the telluric calibrator star observed shortly after we observed the
target.

\subsection{Wavelength Calibration} 

We derived our wavelength calibration based on the location of telluric
absorption features in our spectra.  First, we verified that the wavelength
solution does not change---modulo horizontal offsets due to the variable slit
position--with position on the detector. Then, for each night, we shifted and
stacked all the telluric calibrator spectra using an inverse variance weighted
mean to create a high signal-to-noise template spectrum of the telluric
absorption. 

\begin{figure*}
\epsscale{0.7}
\plottwo{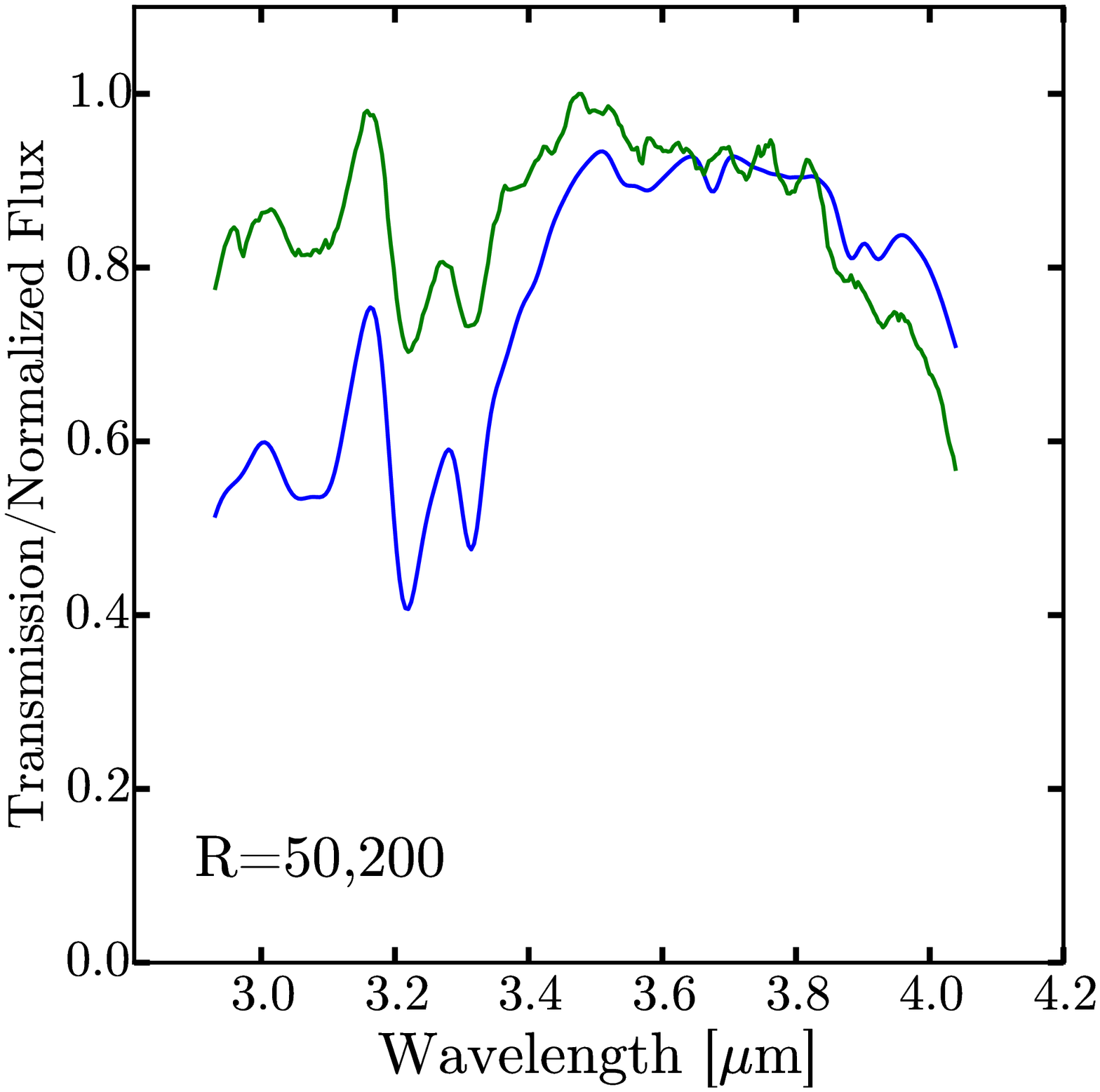}{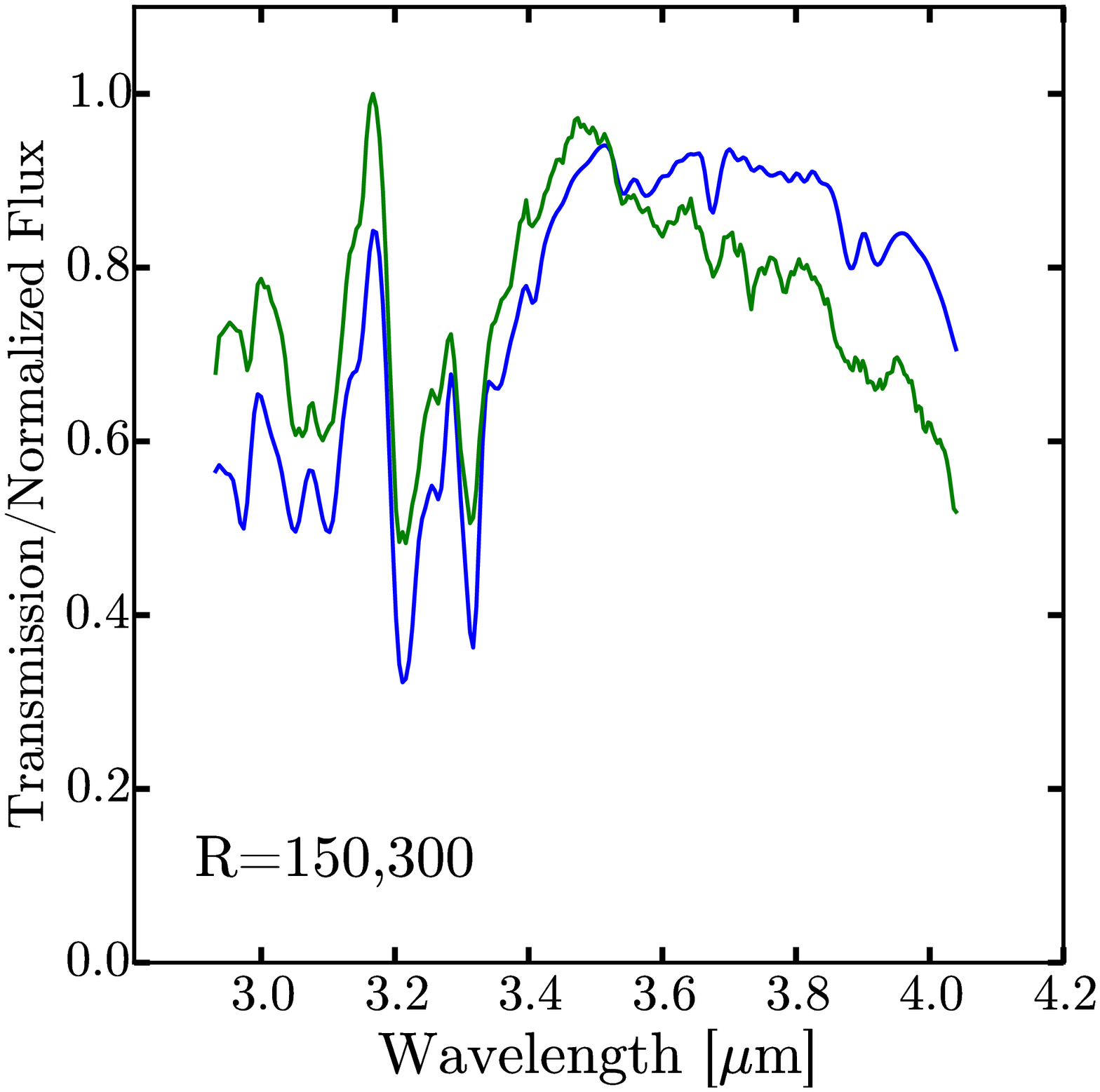}
\caption{Smoothed ATRAN synthetic transmission spectra, demonstrating our
spectral resolution during each run (blue curves). We also show a weighted
average of our observed telluric calibrator spectra (green curves) . The left
side corresponds to our April run, and the right side corresponds to our
November run.\label{resolutionFig}} 
\end{figure*}

To identify telluric absorption features in our stacked telluric spectra, we
used a high-resolution ATRAN \citep{Lord92} synthetic transmission spectrum
tuned for Cerro Pachon and made available by the Gemini
Observatory\footnote{http://www.gemini.edu/sciops/telescopes-and-sites/observing-condition-constraints/ir-transmission-spectra}.
Our approach was iterative, first smoothing the synthetic transmission spectrum
to match a best-guess of the linearly increasing spectral resolution delivered
by the instrument and then comparing the smoothed synthetic spectrum to our
observed spectrum to derive a second-order polynomial transformation to take
pixel position to wavelength. We then performed a two-parameter grid search to
better identify the appropriate linearly increasing resolution parameters for
the synthetic spectrum. This process was repeated until the wavelength solution
and spectral resolution parameters converged. Figure \ref{resolutionFig} shows
our stacked telluric calibrator spectra for each run and the ATRAN telluric
transmission spectrum smoothed using the best-fit linearly increasing
resolution parameters.

The best fit for 2014 April 7 (15.9 mas/pixel mode) was R=50--200. The best fit
for both 2014 November 14 and 15 (25.7 mas/pixel mode) was R=150--300. The
functional forms of the wavelength solutions for each run, modulo horizontal
pixel offsets, are
\begin{equation}
\lambda(x)=(-2.503\times10^{-6}) x^2 + (0.004801) x + 2.822,
\end{equation}
for our April run, with an rms scatter of 0.012 $\mu$m, and
\begin{equation}
\lambda(x)=(-3.634\times10^{-6}) x^{2}  + (0.00548) x + 2.786,
\end{equation}
for our November run, with an rms scatter of 0.007 $\mu$m. We wavelength calibrated the science target spectra by
fitting for the best horizontal offset between target and calibrator, using
a cubic interpolation scheme to perform the shift.

Two of our targets, \mbox{CD-35~2722~B}, and 2M~0103(AB)~b, did not have high enough 
signal-to-noise to provide an adequate fit for the best horizontal offset. For
these sources, we calculated the appropriate horizontal offset using the
measured rotation angle of the slit and the separation from the primary. We
verified using our other targets that this approach yields the correct offset
to within 1--2 pixels.

\subsection{Telluric Calibration}\label{telcal}

Our telluric calibration strategy used primaries to correct companions, and
A-stars to correct primaries. We started by correcting primaries.  We divided
by the A-star spectrum obtained closest in time to a target spectrum and then
multiplied by $\lambda^{-4}$ to correct for the intrinsic black-body shape of
the hot calibrator (this introduces $<2\%$ error). Since most primaries have
a more complicated intrinsic spectrum, before correcting companion spectra we
fit for the best synthetic atmosphere spectrum from the grids of BT-Settl
atmospheres \citep[solar metallicity models from both the 2012 and 2015 grids
were used;][]{Allard12,Baraffe15}. We smoothed the synthetic spectra using the
same linearly increasing resolution parameters that we derived during our
wavelength calibration process. We calibrated companions by dividing by primary
spectra and then multiplying by the best-fit model for the primary. In two
cases, $\eta$~Tel \citep[primary spectral type A0V;][]{Torres06} and \cd
\citep[primary spectral type M1V;][]{Torres06}, we did not obtain an adequate
telluric calibrator spectrum.  In these cases we used spectral types from the
literature to select the appropriate model for the intrinsic shape of the
primary spectrum, using the spectral-type--T$_{\mathrm{eff}}$ relationship of
\citet{Pecaut13}.

\section{Results} \label{ResultsSec}

\begin{figure}
\epsscale{0.7}
\plotone{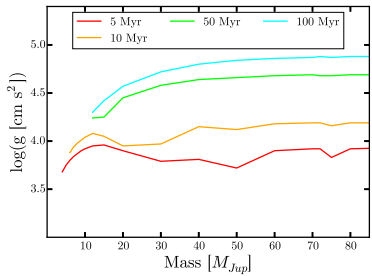}
\caption{Predicted surface gravity as a function of mass from the BT-Settl 2015
models. Each curve represents a different snapshot in age, with the lowest
surface gravities corresponding to 1 Myr, and the highest gravities
corresponding to 100 My. Substellar objects with masses
$\geq2~M_{\mathrm{Jup}}$ and ages older than 1 Myr should have log(g)$>3$.
Likewise objects less massive than 80~$M_{\mathrm{Jup}}$ and younger than 100
Myr should have log(g)$<5$.\label{grav}} 
\end{figure}

\begin{figure*}
\epsscale{0.8}
\plotone{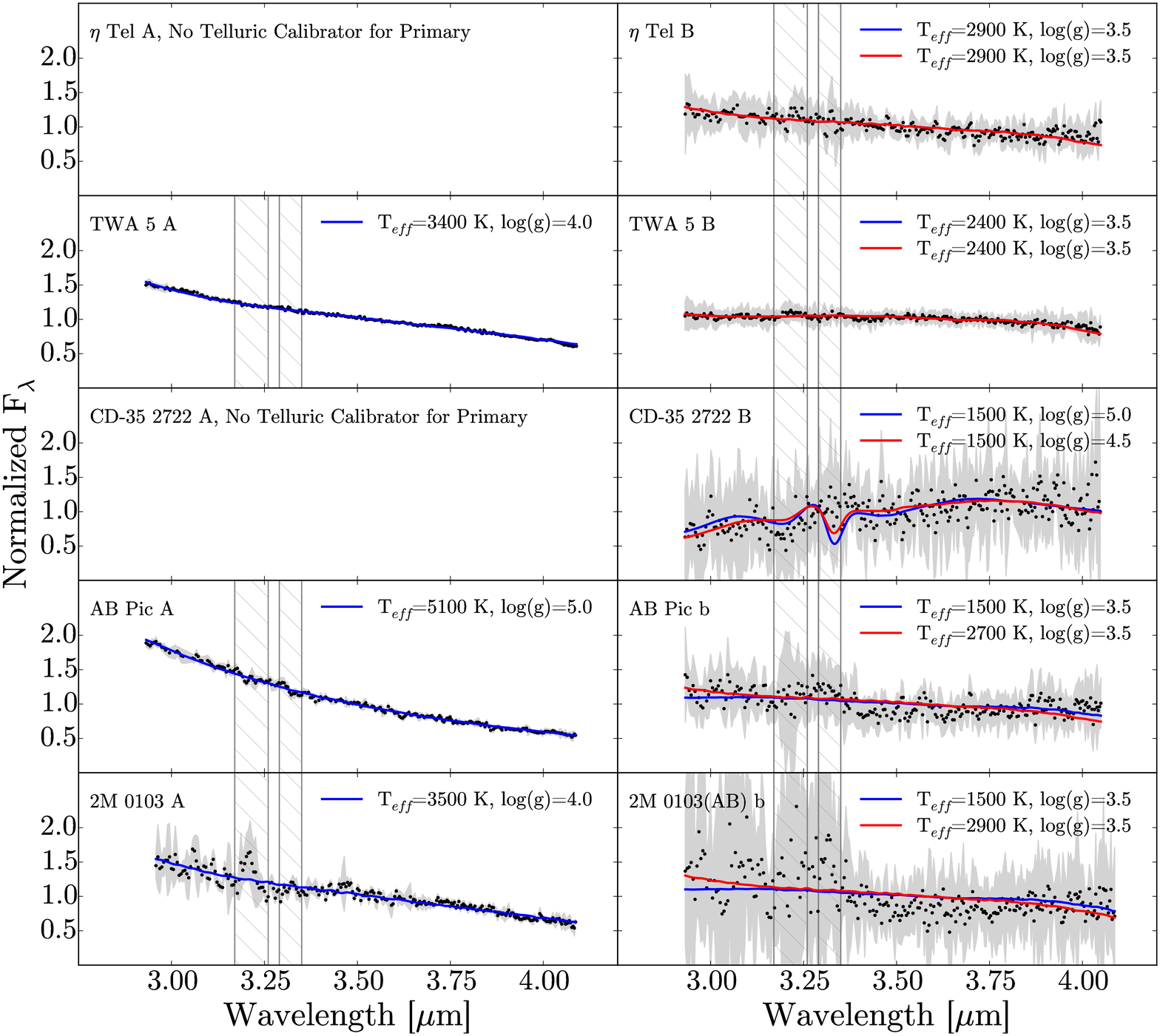}
\caption{L-band spectra of young low-mass companions and their primary stars.
Black points indicate measurements and the gray swath shows the $1-\sigma$
confidence region after scaling bootstrap errors by a factor of 5 to account
for flat-field effects and fixed pattern noise (see Section \ref{fitSec}). We also
show model atmosphere fits to our spectra. In blue, we show the best fit from
the BT-Settl 2015 grid, and in red we show the best fit from the BT-Settl 2012
grid. For our primaries, there were no differences in appearance of best-fit
models from either grid. Hatched regions near 3.2 and
3.3 $\mu$m indicate the location of strong telluric absorption features in our
  data. These regions were not considered during our model fitting. For two of
our primaries, we did not collect appropriate telluric calibrator spectra, so
we do not show their uncorrected spectra. We are still able to correct their
secondaries because we assumed previously reported spectral types to determine
their intrinsic spectral shape.\label{spec}} 
\end{figure*}

\subsection{Model Atmosphere Fits}\label{fitSec}

For each object we determined the best fit model atmosphere from the latest
BT-Settl grid \citep{Baraffe15}. Many of our companions have had BT-Settl 2012
models \citep{Allard12} fit to their near-IR spectra, so we also fit to that
grid to facilitate a direct comparison.  We used models with effective
temperatures ranging from 1200 K to 7000 K, restricting ourselves to models
with solar metallicity and with surface gravity between log(g)=3.5 and
log(g)=5.0.  Figure \ref{grav} illustrates how we chose this range of
gravities.  At ages greater than 5 Myr, even 2~$M_{\mathrm{Jup}}$ objects
should have surface gravity greater than log(g)=3.5. Likewise, even 100 Myr old
brown dwarfs at 80~$M_{\mathrm{Jup}}$ should have surface gravity less than
log(g)=5.0. These restrictions fold in our prior knowledge on the age of our
targets into the fitting process.  After smoothing model spectra to the
resolution of Clio2, fitting was done via $\chi^2$ minimization using the
inverse of our bootstrap errorbars as weights. Our fits ignored the spectral
regions between 3.17--3.26~$\mu$m and 3.29--3.35~$\mu$m because these regions
are affected by strong telluric absorption and are most susceptible to
calibration errors.  The best fits for each object are shown in Figure
\ref{spec}.

The range of allowed model atmosphere parameters consistent with our data
depends on an accurate estimation of our observational uncertainty. As
discussed in Section \ref{extractSec}, we measured the magnitude of random
errors in our data using the bootstrap method, but fixed-pattern noise---from
flat field effects and bad pixels---adds systematic error to our spectra. To
account for this, we scaled up the size of our errorbars using a single
empirically determined scale factor for every spectrum. This scale factor was
determined as follows. We selected our highest signal-to-noise spectra,
excluding the unresolved binary sources TWA~5~A and 2M0103AB.  This left us
with AB Pic A, $\eta$ Tel B, and TWA 5 B. We then scaled the size of the
errorbars for each source until the best-fit model atmosphere provided
a reduced $\chi^2$ of 1. We found scale factors ranging from 3.8 to 5. We
adopted the value of 5 corresponding to the scale factor for AB Pic A because
it is more conservative (larger errors lead to larger allowed ranges for model
parameters) and because it corresponds to a hotter, more massive star where
atmospheric model parameters are more mature and better tested. The size of our
scaled uncertainties are indicated with gray swaths in Figure \ref{spec}.

\begin{figure*}
\epsscale{0.75}
\plottwo{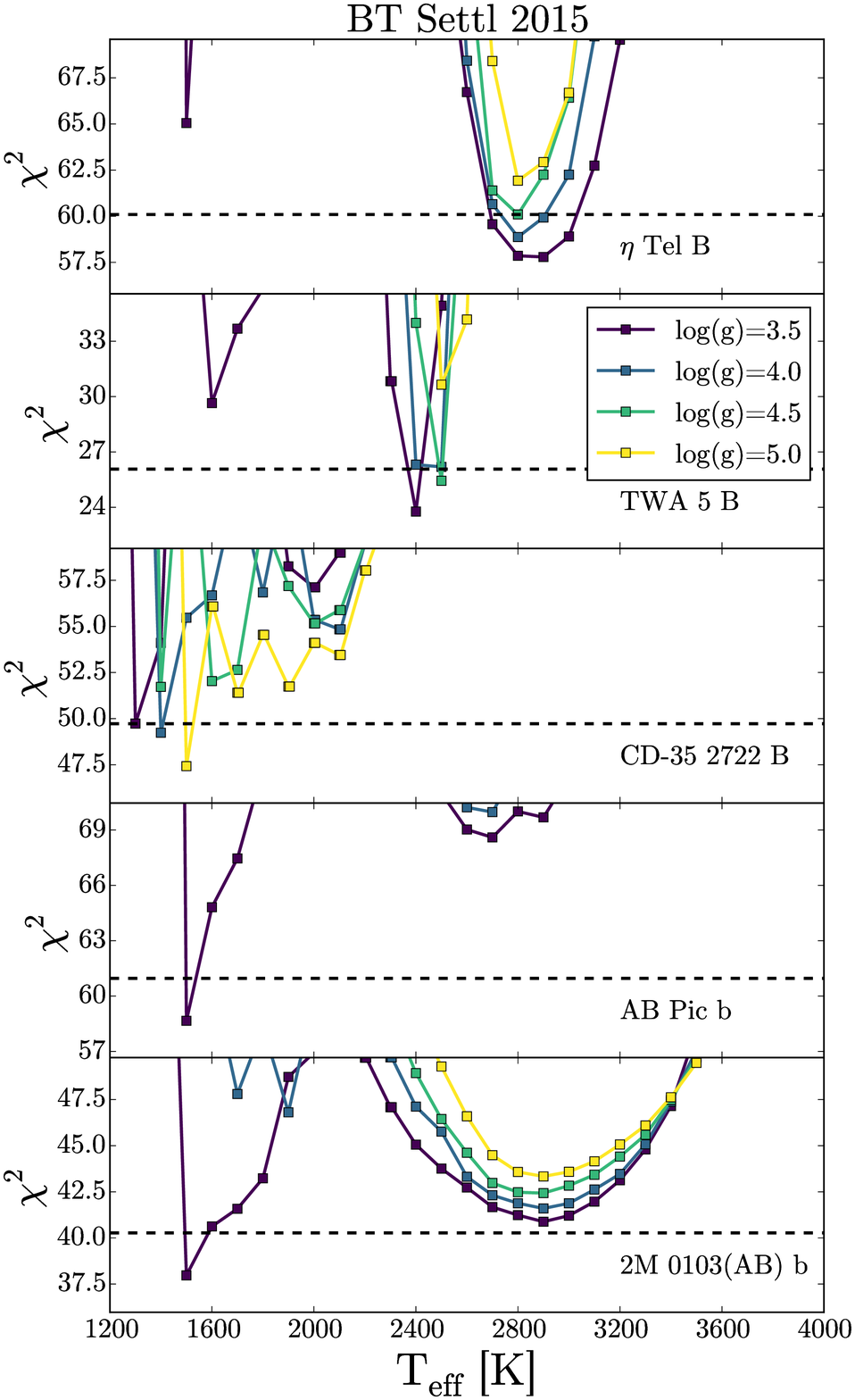}{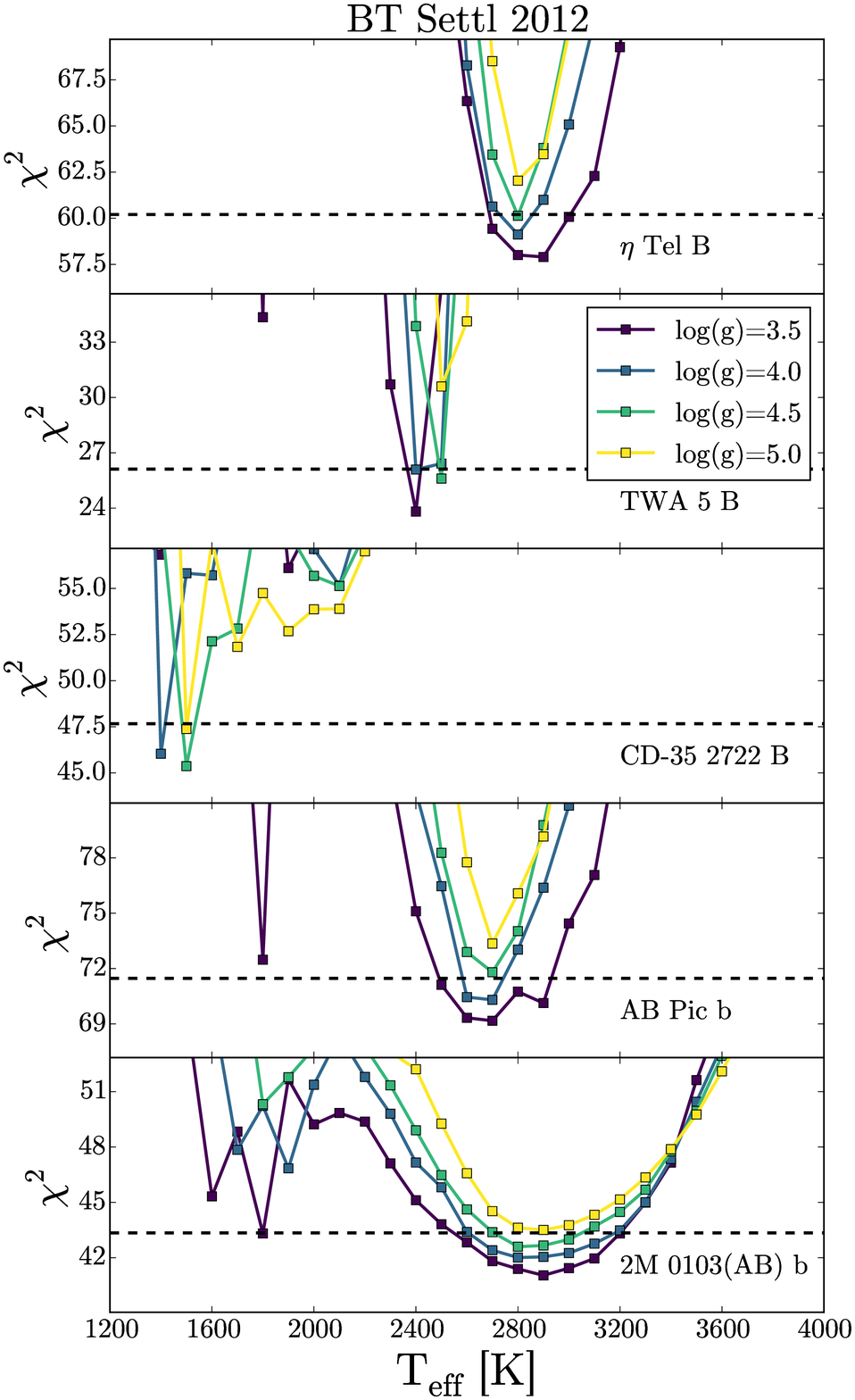}
\caption{Effective temperature versus $\chi^2$ for each of our
secondary spectra. Models with different log(g) have been plotted with
different colors. The best fitting models are shown in Figure \ref{spec}. 
The left panel shows the fitting results for the BT Settl 2015 grid and the
right panel shows the results for the BT Settl 2012 grid. The vertical extent
of each plot reaches $\Delta\chi^2=11.8$: the 99.7\% confidence interval in the
case of gaussian errors. The horizontal dashed line in each panel shows
$\Delta\chi^2=2.3$, the 68\% confidence interval.\label{chiFigs}} 
\end{figure*}

In Figure \ref{chiFigs}, we show $\chi^2$ as a function of temperature and
gravity for each of our sources.  The vertical extent of each plot shows
$\Delta\chi^2=11.8$ above the minimum---this is the 99.7\% (3-$\sigma$)
likelihood interval for the case of gaussian distributed noise. We also
indicate the 68\% (1-$\sigma$) confidence level with a dshed line.

Table \ref{fittable} lists the results of our model atmosphere fitting. We indicate
both the best fit-model atmosphere parameters taken from the grid, as well as the weighted
mean atmospheric parameters calculated according to
\begin{equation}
\bar{m}=\frac{\sum_{i}W_{i}m_{i}}{\sum_{i}W_{i}},
\end{equation}
where the weight $W_{i}$ for each model $m_{i}$ is given by 
\begin{equation}
W_{i}=e^{-0.5\chi^{2}}.
\end{equation}
We followed \citet{Burgasser2010} in calculating sided variance estimates for
our parameters, using
\begin{equation}
\sigma_{m\pm}=\frac{\sum_{i\pm}W_{i}(m_{i}-\bar{m})^{2}}{\sum_{i\pm}W_{i}},
\end{equation}
where the sum is calculated using only parameters above (+) or below (-) the
mean values. For AB~Pic~b and 2M01013(AB)~b the $\chi^2$ curves in Figure
\ref{chiFigs} show two minima with similar values. This will result in very
large weighted variances.

The 2015 and 2012 grids provide mostly similar fits to our spectra, and these
fits reveal atmospheric parameters consistent with those deduced by fits to
1--2.5 $\mu$m spectra. Two of our sources, AB~Pic~b and 2M0103(AB)~b, are fit
by significantly different parameters when using the 2015 and 2012 grid. Figure
\ref{chiFigs} shows that the difference is due to a change in the shape of
model spectra $\sim1500~K$. This can be seen by noting that $\chi^2$ values for
the hotter best-fit models from the 2012 grid remain unchanged in the 2015
grid, yet significantly better fits are obtained at 1500~K.  Both objects
exhibit a jump in the observed flux level near 3.4~$\mu$m that persists to the
blue-end of the spectrum, resulting in a generally blue slope. We discuss these
two targets in more detail in Section \ref{ABsec}.

Most of our best-fit surface gravities appear at the edge of our restricted
range. However, in each case the 1-$\sigma$ confidence interval includes the
next grid value, except for the 2015-grid fits to AB~Pic~b and 2M0103(AB)~b. As
discussed above, these sources exhibit an odd feature that is not predicted by
the models. Where model fits prefer surface gravities at the edge of our grid,
weighted means for this parameter are biased because of the absence of
tests performed outside the grid. Thus we report only limits for this
parameter for some sources in Table \ref{fittable}.

\begin{deluxetable*}{lllllllllll}
\tabletypesize{\footnotesize}
\tablecolumns{11}
\tablewidth{0pt}
\tablecaption{BT-Settl Model Atmosphere Fits to L-band Spectra}
\tablehead{
\colhead{}&
\multicolumn{4}{c}{BT-Settl 2015}&
\multicolumn{4}{c}{BT-Settl 2012}&
\multicolumn{2}{c}{Fits to NIR spectra\tna}\\
\colhead{Target} &
\colhead{$T_{\mathrm{eff,grid}}$} &
\colhead{log(g)$_{\mathrm{grid}}$}&
\colhead{$T_{\mathrm{eff,mean}}$} &
\colhead{log(g)$_{\mathrm{mean}}$}&
\colhead{$T_{\mathrm{eff,grid}}$} &
\colhead{log(g)$_{\mathrm{grid}}$}&
\colhead{$T_{\mathrm{eff,mean}}$} &
\colhead{log(g)$_{\mathrm{mean}}$}&
\colhead{$T_{\mathrm{eff}}$} &
\colhead{log(g)}
}
\startdata
$\eta$~Tel~B  &2900 & 3.5 & $2838^{+114}_{-153}$          & $<4.6$             & 
    2900& 3.5& $2834_{-80}^{+115}$   & $<4.6$              & 
        2600$\pm$100        &4$\pm$0.5\\
TWA~5~B       &2400 & 3.5 & $2410_{-168}^{+90}$        & $<4$             & 
    2400& 3.5& $2431_{-54}^{+70}$   & $<4$              & 
        2500$\pm$100        &4$\pm$0.5\\
\cdB          &1500 & 5.0 & $1596_{-176}^{+311}$& $>3.9$& 
    1500& 4.5& $1497_{-98}^{+136}$   & $4.4_{-0.5}^{+0.3}$ & 
        1700-1900           &4.5$\pm$0.5\\
AB~Pic~b      &1500 & 3.5 & $1546_{-46}^{+720}$ & $<4.3$             & 
    2600\tnb& 3.5& $2664_{-271}^{+136}$ & $<4.1$              & 
        1800$^{+100}_{-200}$&4.5$\pm$0.5\\
2M~0103(AB)~b &1500 & 3.5 & $2443_{-592}^{+794}$& $<4.4$            & 
    2900\tnb& 3.5& $2858_{-319}^{+189}$& $<4.4$               & 
        ---                 &---   
\enddata
\tablenotetext{a}{Same as in Table \ref{targSumTable}, included here for convenience.}
\tablenotetext{b}{The $\chi^2$ surfaces for AB~Pic~b and 2M~0103(AB)~b show two
minima with one preferred by fits to the 2012 grid and the other preferred by
the 2015 grid. Photometry indicates that the lower-temperature fits are more
accurate (see Section \ref{ABsec}).}
\label{fittable}
\end{deluxetable*}
\subsection{Comparison to Field Dwarf Spectra}
\label{fieldsec}
\begin{figure*}
\epsscale{0.8}
\plottwo{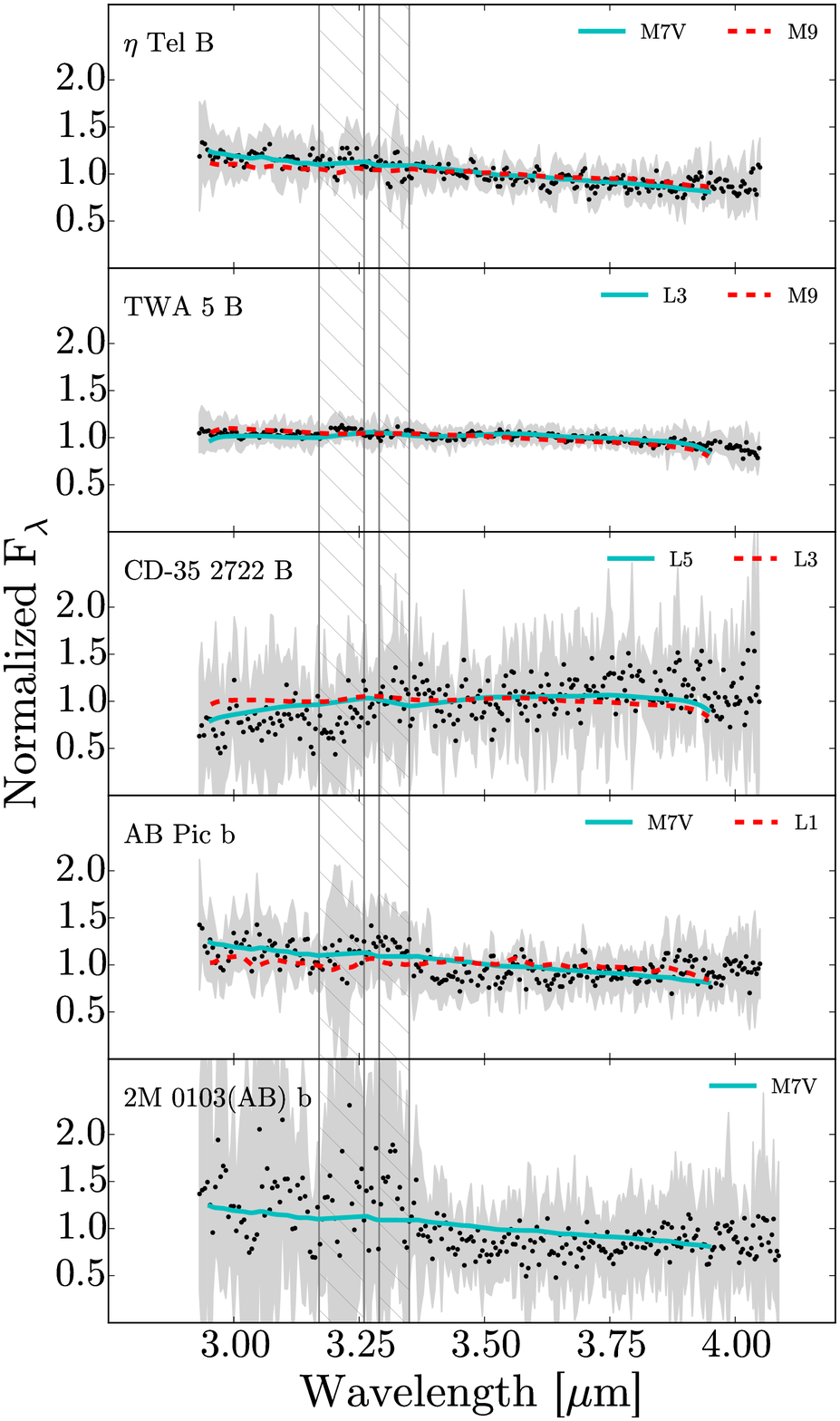}{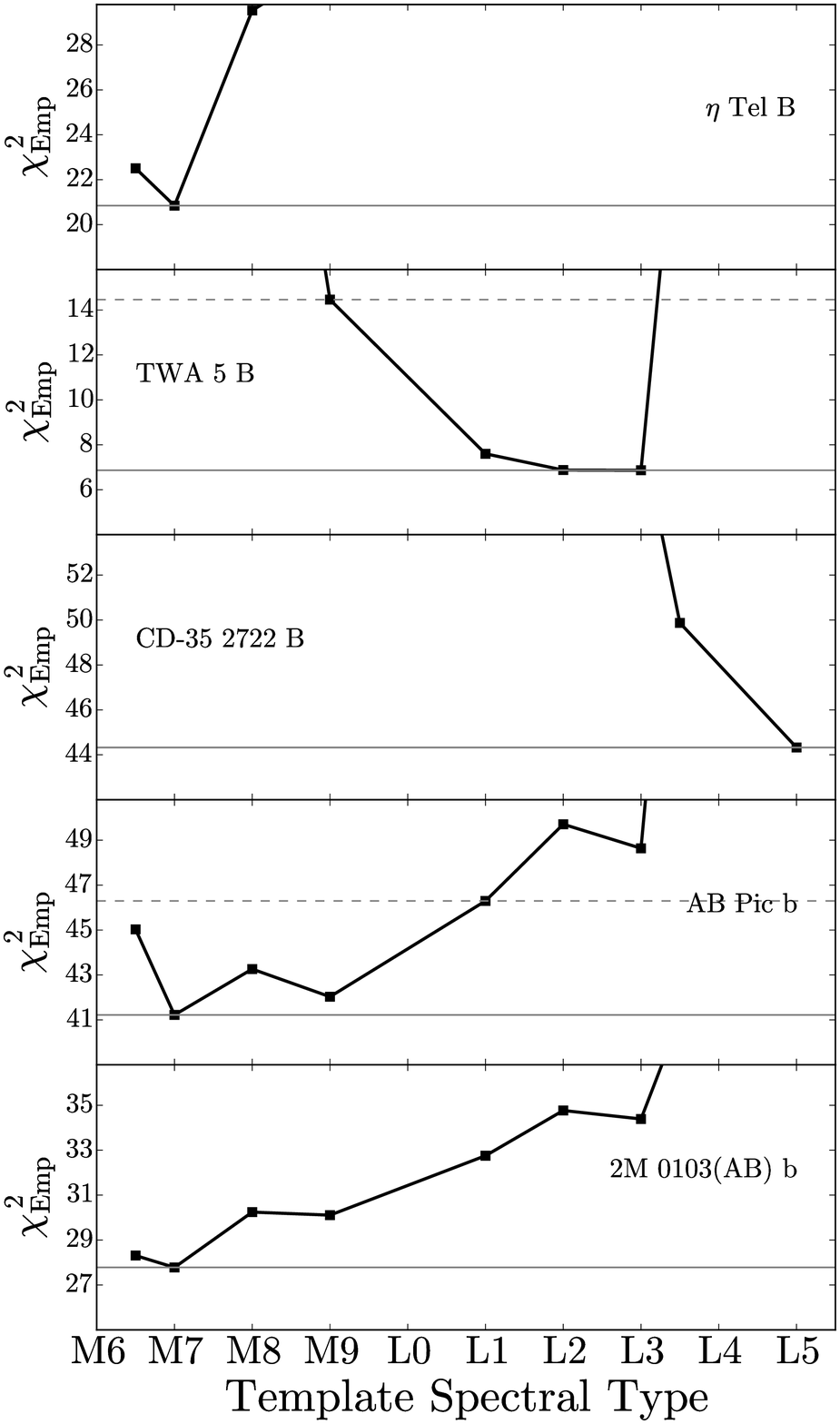}
\caption{Left: The same companion spectra as in Figure \ref{spec}, now compared to
empirical spectra of field dwarfs from \citet{Cushing05}. With a solid cyan
line we show the best fit field dwarf spectrum. We show the spectrum
corresponding to the optical/NIR spectral type of each of our targets (Table
\ref{targSumTable}) with a red dashed line.  The field spectra for each spectral type
correspond to the following targets: M7, GJ~644~C; M9, LP~944-20; L1,
2MASS~J14392836+1929149; L3, 2MASS~J15065441+1321060; L4.5,
2MASS~J22244381-0158521; L5, 2MASS~J15074769-1627386. Right: Spectral type
versus $\chi^2$ resulting from our fitting analysis (solid black curves). The
vertical extent of each plot shows the 99.7\% confidence interval in
$\Delta\chi^2$. We highlight the best fit model with a horizontal solid gray
line. We also indicate the chi-square value of optical/NIR spectral type with
a dashed line.\label{spec2}} \end{figure*}

The L-band SEDs of directly imaged extrasolar planets appear distinct from the
SEDs of older field dwarfs at the same effective temperatures \citep{Skemer14}.
The discrepancy needs better characterization to provide insight into the
physical processes affecting their atmospheres. Comparison to model atmospheres
can be helpful in identifying important spectral features, but models are
challenged to capture the processes at play in cool atmospheres, particularly
in providing accurate representation of clouds \citep[e.g.,][]{Marley14}.  Here
we compare our spectra of young directly imaged companions to the spectra of
field dwarfs from the literature. We use the spectra of M7 through L5 type
objects presented in \citet{Cushing05}. We smoothed the template spectra to the
resolution of our data and interpolated them to sample the same wavelengths. We
then performed the fit by minimizing
\begin{equation}
\chi^{2}_{\mathrm{Emp}} = \sum_{\lambda}
\frac{(S_{\lambda,\mathrm{Clio2}}-S_{\lambda,\mathrm{Spex}})^{2}}{\sigma_{\lambda,\mathrm{Clio2}}^{2}
+ \sigma_{\lambda,\mathrm{Spex}}^{2}},
\end{equation}
where $S_{\lambda,\mathrm{Clio2}}$ is our MagAO/Clio2 spectrum and
$S_{\lambda,\mathrm{Spex}}$ is the smoothed interpolated field dwarf spectrum
from \citet{Cushing05}, and the $\sigma$'s are the associated errors for each
spectrum. We used the same errors for all the field dwarf template spectra,
which were determined by requiring that two examples of M9-type spectra from
\citet{Cushing05} fit each other well ($\chi^{2}_{\mathrm{Emp}}=1$). This helps
incorporate the uncertainty in spectral type for our template spectra into our
fitting analysis.

Figure \ref{spec2} shows the best fitting field dwarf spectrum for each of our
companion sources.  We also overplot the field dwarf spectrum corresponding to
the optical/NIR spectral type reported for each of our sources (or the closest
sub-type available), except 2M~0103(AB)~b, which does not have a NIR spectral
type reported in the literature.  

In general, the best-fit spectrum and the spectrum selected to match the
optical/NIR spectral type are consistent---especially when accounting for the
uncertainty in the spectral classification both for our sources and for the
field objects.  Our data show that for late-M and early L type spectra, the
L-band is not particularly diagnostic of spectral type. For example, the M9 and
L3 type dwarf spectra both track the spectrum of TWA~5~B and these are formally
both allowed by the data as shown in the right panel of Figure \ref{spec2},
where we quantify the similarities among the fits to different spectral types
by showing spectral type versus $\chi^{2}_{\mathrm{Emp}}$ for each of our
targets.  However, the formal $\Delta\chi^2$ analysis should be viewed with
caution given the strong influence of systematic uncertainty both in our data
and the Spex data.  Previous authors have also noted that L-band spectra alone
are not particularly powerful for deducing spectral types in this range
\citep[e.g.,][]{Cushing08}.

At later spectral types 3--4~$\mu$m spectra do exhibit distinct features
compared to late M-dwarfs and early L-dwarfs because of increased methane
opacity and decreased cloud opacity at cooler temperatures. The latest spectral
type source in our sample, CD-35~2722~B, which is classified as L3 based on NIR
spectral indicators \citep{Allers13}, appears redder than the best fit L5 field
dwarf spectrum, with most of our measurements below the dwarf spectrum at the
blue end of the band, and most of our measurements above the dwarf spectrum at
the red end.  This example suggests the importance of L-band spectroscopy for
objects with later spectral type than mid-L, where the first hints of CH$_{4}$
absorption are expected to arise \citep{Noll2000}.

\section{Discussion} \label{DiscSec}
\subsection{Comparison to Directly Imaged Planets}
The low surface gravity prevalent in young planetary mass atmospheres is
responsible for the presence of clouds and dis-equilibrium chemistry at
effective temperatures $\sim800$ to $1300$~K, where older more massive T-dwarfs
appear cloud free and methane rich \citep{Marley1996,Barman11b,Skemer12,Marley12}. Here we
observe a warmer more-massive population of young companions.  Most of our
atmospheric fits are consistent with the fits found by previous authors using
NIR spectra.  The similarity of our L-band derived atmospheric parameters
compared to near-IR results demonstrates the functionality of the MagAO/Clio2
spectroscopic mode and the general success of model atmospheres to capture the
most important physics responsible for the appearance of spectra at these
effective temperatures, where clouds are expected and CH$_{4}$ absorption is
not.

Our data allow us to constrain the onset of peculiarities where young
low-gravity objects appear to have L-band spectra distinct from field objects.
This range must be below $\sim1500$~K, the effective temperature found here for
AB~Pic~b and 2M~0103(AB)~b. This interval of effective temperatures
includes the L-T transition for field brown dwarfs, emphasizing that the
discrepancies in 3--4~$\mu$m spectra may be related to the cloud clearing and
chemical processes responsible for the evolution from L-dwarf to T-dwarf.

\subsection{\cdB} 

\citet{Allers13} assigned a spectral type of L3 with intermediate surface gravity
for \cdB based on an average of several indicators. Those authors noted \cdB
exhibited significantly higher gravity than 2MASS~J03552337+1133437, which has
similar infrared spectral type and age \citep{Liu13B}. Among our targets, fits
to BT-Settl models suggest that \cdB exhibits the highest surface gravity,
consistent with previous studies and the relatively old age of the object
compared to the rest of our sample.

\cdB was classified as an L4 by \citet{Wahhaj11} based on the similarity of its
J, H, and K-band spectra to 2MASS~J2224438-015852, an object whose very red
color is distinct from other field dwarfs and implies large amounts of
atmospheric dust \citep{Cushing05,Stephens09}. \cdB has a bluer NIR color than
2MASS~J2224438-015852 \citep{Wahhaj11}, suggesting somewhat different cloud
structure.

Our spectrum of \cdB is redder than the L5 field dwarf template spectrum we
compared it to and best-fit model atmospheres are cooler than reported for NIR
fits. These results are driven by a blue slope to our spectrum. Since we do not
have a telluric calibrator for the primary star in the \cd system, we assumed
the intrinsic shape of the star (based on its spectral type) when we calibrated
the substellar companion. To check whether the blue slope of \cdB and our
fitted parameters for this object could be due to selecting an overly red
intrinsic spectrum for the primary, we repeated our telluric calibration
procedure using a model corresponding to two spectral subtypes earlier than
originally used. This did not result in a hotter best-fit model atmosphere.

\cdB is the oldest target in our young sample \citep[$\sim50$
Myr,][]{Zuckerman04}, and has the highest surface-gravity. If its red
3--4~$\mu$m spectral slope does indicate thinning clouds, our observations
suggest that this source may be cooler and of later type than previously
thought, but this would need to be confirmed with higher signal-to-noise
observations.

\subsection{AB~Pic~b and 2M~0103(AB)~b}\label{ABsec}

As we pointed out in Section \ref{fitSec}, the BT-Settl 2012 and 2015 grids
provide very different best-fit effective temperatures for AB~Pic~b and
2M~0103(AB)~b. Each of these targets is better fit with a 1500~K model from the
2015 grid while fits to the 2012 grid show 2600~K and 2900~K models are
     preferred for AB~Pic~b and 2M0103(AB)~b, respectively. However, models
with different effective temperatures have different predicted total fluxes.
For an age of 30 Myr, the BT-Settl evolutionary models predicts L' absolute
magnitudes of 8.7 mags and 7.5 mags for 2600~K and 2900~K effective
temperatures, respectively \citep{Baraffe15}. These are both significantly
brighter than the absolute L' magnitudes reported by \citet{Delorme13} for
AB~Pic~b ($9.9\pm0.1$) and 2M0103(AB)~b ($9.5\pm0.1$). According to the models,
the observed fluxes correspond to objects with effective temperatues $<
1700~K$.

Both AB~Pic~b and 2M~0103(AB)~b exhibit an increase in flux in the spectral
region between $\sim2.9$--$3.4$~$\mu$m.  While part of this range includes
a region of strong telluric absorption which challenges precise calibration,
a significant portion is outside the strongest telluric features. None of the
atmospheric models predict such a feature and the high effective temperatures
suggested by the BT-Settl 2012 fits are probably incorrect, caused by the
average blue slope of the spectra due to the anomalous blue feature.

For 2M~0103(AB)~b, the feature may be due to a combination of lower
signal-to-noise and difficulties in observing and calibrating associated with
orbiting a close binary (e.g., reduced AO performance). However, for AB~Pic~b,
we achieved higher signal-to noise across the spectrum.  The spectrum of
AB~Pic~b was corrected for telluric emission using the spectrum of AB~Pic~A,
which was positioned within the slit with AB~Pic~b and observed simultaneously;
any atmospheric changes during the course of the observation should have been
tracked perfectly by the primary.  Furthermore, our model atmosphere fit for
AB~Pic~A is exactly as expected for the spectral type \citep[K1V;][]{Torres06,
Pecaut13}, suggesting we have properly calibrated AB~Pic~A and identified the
correct model for its intrinsic spectral shape when correcting AB~Pic~b. Given
the quality of the telluric correction of AB~Pic~A, we also checked whether
correcting AB~Pic~b with the same A-star calibrator provided a different
result. Regardless of whether HD~32007 or AB~Pic~A is used for telluric
calibration, the increased flux short of 3.4 $\mu$m persists.  We also verified
that the feature is present in both A-nods and B-nods independently and in the
first half of our integrations as well as in the second half. 

If the increase in our spectrum is physical, the larger-than-expected flux in
the spectral region corresponding to methane opacity could indicate an
atmospheric inversion or possibly an aurora \citep[e.g.;][]{Hallinan15}.
However, strong H$_{3}^{+}$ emission, often associated with aurorae
\citep[e.g.,][]{Brown03}, is not present at 3.5 and 3.7 $\mu$m.  Polycyclic
aromatic hydrocarbon molecules (PAHs) are known to have broad emission bands
centered at 3.3 and 3.4 microns. However, fluorescent PAHs in the vicinity of
AB~Pic~b and 2M~0103(AB)~b are unlikely given the system ages.  While AB~Pic~A
is known as a faint EUV and X-ray source \citep{Lampton1997}, this seems
unlikely to cause strong PAH excitation. Additional follow up
observations of AB~Pic~b in the L-band are warranted to better understand the
blue feature seen in our data.

\section{Conclusions} 

We commissioned the L-band spectroscopic mode of Clio2 behind the MagAO system.
We observed five young systems with directly imaged low-mass companions. The
recovered spectra of primary stars match templates for their literature
spectral types, confirming the fidelity of the observations. The spectra of
companions are consistent with expectations based on fits to shorter wavelength
data. This result constrains the temperature range where the L-band SEDs of
young directly imaged planets begin to diverge from older field dwarfs to
$T_{\mathrm{eff}} \lesssim 1500$~K. This range includes the
the L-T transition for field dwarfs, providing further evidence that observed
discrepancies are due to clouds and non-equilibrium CO/CH$_{4}$ chemistry due
to vertical mixing. The L-band spectrum of \cdB is redder than field dwarfs and
suggests some cloud settling in this $\sim50$~Myr old object but low signal to
noise limits our ability to draw any firm conclusions.  We also see an
increased flux feature from 2.95--3.4~$\mu$m in the spectra of 2M~0103(AB)~b
and AB~Pic~b.  The feature coincides with methane bandheads, and if
physical, could indicate a thermal inversion.  L-band spectroscopy with Clio2
will be an important complement to ongoing surveys for directly imaged
extrasolar planets that are conducting their searches at shorter wavelengths.

\acknowledgements{ This work was supported by NASA Origins grant NNX11AK57G and
NSF AAG grant 121329. J.M.S. was also partially supported by the state of
Arizona Technology Research Initiative Fund Imaging Fellowship. A.S. and T.J.R.
are supported by the National Aeronautics and Space Administration through
Hubble Fellowship grants HST-HF2-51349 and HST-HF2-51366.001-A (respectively)
awarded by the Space Telescope Science Institute, which is operated by the
Association of Universities for Research in Astronomy, Inc., for NASA, under
contract NAS 5-26555. K.M.M. was supported under contract with the California
Institute of Technology, funded by NASA through the Sagan Fellowship Program.
We thank M. Cushing for providing L-band spectra of field
dwarfs.}

\end{document}